\documentstyle[eqsecnum,aps,prd,epsf,preprint]{revtex}
\newcommand{\snab}{\nabla \hspace{-3.7mm}{\not}\hspace{3.7mm}}  
\newcommand{\spar}{\partial \hspace{-2.9mm}{\not}\hspace{2.9mm}}

\newcommand{\barh}{{h} \hspace{-2.4mm}{{}^-}\hspace{2.4mm}}

\begin{document}
\draft

\title{
Explicit Derivation of the Fluctuation-Dissipation Relation \\
of the Vacuum Noise \\
in the $N$-dimensional de Sitter Spacetime}

\author{T.~Murata, K.~Tsunoda \and K.~Yamamoto}

\address{Department of Physics, Hiroshima University,
         Higashi-Hiroshima 739-8526, Japan}

\date{Jan. 2001}
\maketitle

\begin{abstract}
Motivated by a recent work by Terashima (Phys.~Rev.~D60 084001), 
we revisit the fluctuation-dissipation (FD) relation between the 
dissipative coefficient of a detector and the vacuum noise of fields
in curved spacetime. In an explicit manner we show that the 
dissipative coefficient obtained from classical equations 
of motion of the detector and the scalar (or Dirac) field 
satisfies the FD relation associated with the vacuum noise of 
the field, which demonstrates that the Terashima's 
prescription works properly in the $N$-dimensional de Sitter spacetime. 
This practice is useful not only to reconfirm the validity of 
the use of the retarded Green function to evaluate 
the dissipative coefficient from the classical equations of motion 
but also to understand why the derivation works properly, which 
is discussed in connection with previous investigations on the 
basis on the Kubo-Martin-Schwinger (KMS) condition. 
Possible application to black hole spacetime is also briefly 
discussed.
\end{abstract}

\vspace{2mm}
\section{introduction}
It is well-known that the vacuum fluctuations of the Minkowski 
spacetime can be a thermal ambience when viewed by uniformly 
accelerated observers. 
This effect, which is called the Unruh effect, has been 
investigated by many authors.
Recently Terashima \cite{ref-1} has proposed a new method 
to evaluate the Rindler noise indirectly by using the 
fluctuation-dissipation (FD) relation (or theorem) formulated by 
Callen and Welton \cite{ref-3}. He has explicitly shown that 
the FD relation holds between the Rindler noise and the 
dissipative coefficient of the De Witt detector which is 
obtained from classical equations of motion of the 
detector and the (scalar and Dirac) fields in $N$-dimensional 
Rindler spacetime.

The FD relation is a general relation between the 
spontaneous fluctuations of generalized forces in thermal 
equilibrium and dissipation in linear dissipative system. 
Studies on the FD relation has a long history, and provide 
a basis of statistical physics for dissipative systems 
(e.g., \cite{ref-2,ref-6} and references therein). 
The Callen and Welton's formula \cite{ref-3} is one of the general 
formulations of the FD relation developed in a system of many 
microscopic degrees of freedom in the thermal equilibrium state 
and a macroscopic degree of freedom, which are linearly coupled 
with each other. 

Though several authors have investigated the vacuum noise of 
quantum fields in curved spacetime from a view point of the 
FD relation \cite{ref-4,Ooguri,CS,SCD}, however, Terashima's 
approach is slightly different from those previous approaches. 
Following the Terashima's prescription, the dissipative coefficient 
of the detector's inner motion is explicitly calculated from the classical 
equations of motion of the detector and the fields. 
Then the Rindler noise is simply obtained by inserting the 
dissipative coefficient into the FD relation, which relates 
the dissipative coefficient to the noise of vacuum fluctuations. 
He has shown that this method is successful in the system of the 
De Witt detector coupled to the scalar field in $N$-dimensional 
Rindler spacetime, and he has generalized the method to the Dirac 
field by using the fermionic FD relation. 

In two and four dimensional spacetime, it is well-known that the 
spectrum of the Rindler noise is completely equivalent to the thermal 
noise spectrum with the temperature $T=\alpha/(2\pi)$, where 
$\alpha$ is the acceleration of the detector. 
Therefore one might think that the existence of the FD-relation 
is trivial. In general spacetime dimensions, however, there are 
differences between the thermal noise and the Rindler noise, 
which is well recognized by previous investigations. 
For example, Takagi found the 'statistical inversion' phenomenon 
\cite{ref-4}, that is, the Rindler noise spectrum of massless 
particle in odd dimension becomes the Fermi-Dirac distribution 
instead of the Plank one, on the contrary to naive expectation.

In the meanwhile, de Sitter spacetime is one of the most famous 
curved spacetime which has been studied by many theorists, e.g., 
\cite{GH,ref-5}.  
The reason of this comes from the fact that de Sitter spacetime is 
the unique maximally symmetric curved spacetime and also that quantum 
fluctuations in de Sitter spacetime is considered to be the origin 
of the cosmic structure in the inflationary universe scenario.
In the quantum field theory in the de Sitter spacetime, it is well-known 
that the vacuum noise exhibits similar character of the Rindler 
noise including the phenomenon of the statistical inversion.
Motivated by the investigation by Terashima, we apply his method
to the vacuum noise in the de Sitter spacetime.
We will show that the FD relation of Callen and Welton holds for the 
de Sitter noise of the scalar and the Dirac fields in $N$-dimensional de 
Sitter spacetime following the prescription done in the Rindler 
spacetime by Terashima. This practice is useful not only to
understand the reason why Terashima's method works properly 
irrespective of the existence of the statistical inversion 
but also to clarify the relation between the Terashima's 
investigation and previous investigations of the FD relation 
of the vacuum noise in curved spacetime \cite{ref-4,Ooguri,CS,SCD}, 
in which the periodicity of Green functions in imaginary time 
is attributed to the thermal nature. Our investigation is also 
instructive to demonstrate the validity of the use of the 
retarded Green function to evaluate the dissipative coefficient 
from the classical equations of motion, which gives a principle 
to avoid singular poles in the Terashima's prescription.

This paper is organized as follows: In \S 2 we calculate 
the dissipative coefficient of the detector from the classical 
equations of motion of the detector and the fields in $N$-dimensional 
de Sitter spacetime. Then we explicitly show 
that the bosonic and fermionic FD relations exist between the
vacuum noise and the dissipative coefficient. 
In \S 3 we rephrase the calculations in \S 2 from a general 
point of view by summarizing relations that hold between the
Green functions, where key conditions so that the prescription 
of Terashima works properly are made clear. 
\S 4 is devoted to summary and discussions. Throughout this paper we 
use the natural unit $c=\barh=k_{B}=1$, and adopt the convention 
$(1,-1,-1,-1)$ for metric signature and curvature tensor 
and $\{\gamma^{\mu},\gamma^{\nu}\}=2g^{\mu\nu}$ for gamma 
matrices \cite{ref-5}.

\section{Explicit calculation of the FD-relation in de Sitter Spacetime}
\subsection{real scalar field}
\def\x{{x_0}}
\def\L{L}
\def\bfx{{\bf x}}
\def\deltatau{{\Delta\tau}}
 In this section we apply the Terashima's prescription to the 
$N$-dimensional de Sitter spacetime. We first consider a real 
scalar field $\phi$ with mass $m$ and a detector which couples 
linearly with the field with the action :
\begin{equation}
 S=S_{0}(q)+S_{int}(q,\phi)+S_0(\phi),
\label{action}
\end{equation}
where
\begin{eqnarray}
 &&S_{0}(q)=\int d\tau \L(q,\dot{q}) ,
\label{1-1}
\\
 &&S_0(\phi) =\int {d^nx} \sqrt{-g}\   \frac{1}{2} 
     \left\{ g^{\mu\nu} \partial_\mu\phi  \partial_\nu\phi 
                  -(m^2+\xi R)\phi^2 \right\} ,
\label{1-2}
\\
 &&S_{int}(q,\phi) 
  =  \int d\tau {d^nx} q(\tau) \phi(x) \delta^{n}\left(x-\x(\tau)\right)
  = \int d\tau q(\tau) \phi(\x(\tau)) ,
\label{1-3}
\end{eqnarray}
where $\L$ denotes the Lagrangian of the degree of freedom of 
the detector $q(\tau)$, $\tau$ is the proper time, $\x(\tau)$ 
denotes the world line of the detector, $R$ is 
the Ricci scalar, $\xi$ is the non-minimal coupling constant, 
and $\delta^{n}\left(x-y\right)$ is the $n$-dimensional
Dirac delta function. From this action we have equations of motion,
\begin{eqnarray}
 &&\left( \frac{\delta S_0}{\delta q}\right) \left(\tau \right)
   + \int d^n x \delta^n (x-\x(\tau)) \phi (x)=0,
\label{1-4} \\
  &&\left( \nabla^\mu \nabla_\mu +m^2 +\xi R \right) \phi(x) 
   =\frac{1}{\sqrt{-g}} \int d\tau q(\tau) \delta^n (x-\x(\tau)),
\label{1-5}
\end{eqnarray}
where we defined
\begin{eqnarray}
 \left(\frac{\delta S_0}{\delta q} \right)(\tau)\equiv 
      - \frac{d}{d\tau} \left( \frac{\partial \L}{\partial \dot{q}}\right) 
         +\frac{\partial \L}{\partial q}.
\end{eqnarray}

The line element of the $N$-dimensional de Sitter spacetime
can be written as 
\begin{eqnarray}
 {ds}^2 = {dt}^2 - e^{2H t} \sum^{n-1}_{m=1} ({dx^m})^2
  =(-H \eta)^{-2} \eta_{\mu \nu} {dx}^\mu {dx}^\nu,
\end{eqnarray}
where $H$ is the Hubble parameter, $\eta_{\mu \nu}$ is the metric 
of the flat spacetime, and we introduced the conformal time 
$\eta$ in the last equality, which is related to the cosmic 
time $t$ as 
\begin{equation}
 \eta \equiv - \frac{1}{H} e^{-H t}\mbox{ }
  ,\mbox{ }~~~~~~~~~\left( -\infty < \eta < 0 \right).
\label{1-6}
\end{equation}
For simplicity let us consider the massless field with the conformal 
coupling and the detector at rest of the coordinate 
$(t=\tau,~\bfx=\bfx_0)$, where 
$\bfx=(x^1,x^2,\cdots,x^{n-1})$. In this case equations of motion 
$(\ref{1-4})$ and $(\ref{1-5})$ can be rewritten as
\begin{eqnarray}
 &&\left( {\frac{\delta S_0}{\delta q}}\right) \left( \tau\right) 
  + (-H\eta(\tau))^{n/2 -1} \bar{\phi}(\x) =0,
\label{1-7} \\
 &&\eta^{\mu \nu} \partial_\mu \partial_\nu \bar{\phi}
  =  (-H\eta )^{n/2-2} \delta^{n-1} (\bfx-\bfx_0) q(t(\eta)),
\label{1-8}
\end{eqnarray}
by introducing the conformal field $\bar{\phi}$, which is
defined with a conformal weight as
\begin{eqnarray}
 \phi = (-H\eta )^{n/2 -1} \bar{\phi}.
\end{eqnarray} 

Next we consider equation (\ref{1-8}). According to the Terashima's
method, equations for Fourier coefficients are first derived,
then the equations are solved. To obtain the dissipative 
coefficient, he used an $i\epsilon$-prescription. Namely, 
imaginary part arising from this $i\epsilon$ prescription 
leads to the dissipative coefficient. To solve equation 
$(\ref{1-8})$, we here adopt the Green function method. 
It is natural to adopt the retarded Green 
function when treating classical equations of motion. 
Terashima's method is equivalent to adopting this Green function.
The retarded Green function satisfies the equation
\begin{equation}
  \biggl(\frac{\partial^2}{\partial \eta^2}-
  \frac{\partial^2}{\partial \bf x^2}\biggr)
  {\bar G}_R (x,x^\prime)
         =\delta^n (x-x^\prime),
\label{1-9}
\end{equation} 
and the boundary condition ${\bar G}_R(x,x^\prime)=0$ 
for $\eta-\eta^{\prime}< 0$.
Then equation $(\ref{1-8})$ can be solved, and $\bar{\phi}(x)$ 
can be written, 
\begin{eqnarray}
 \bar{\phi}(x) & = & \int_{-\infty}^0 d\eta^\prime \int_{-\infty}^\infty
     d^{n-1}\bfx^\prime {\bar G}_R(x,x^\prime) 
     \left(-H \eta^\prime \right)^{n/2-2}
      \delta^{n-1}(\bfx^\prime-\bfx_0) q(t'(\eta'))
\nonumber \\
 & = & -\int_{-\infty}^0 d\eta^\prime 
   \int_{-\infty}^\infty \frac{d^n p}{(2\pi)^n} 
  \frac{e^{ip^0(\eta-\eta^\prime)}}{p^2}
     \left(-H \eta^\prime \right)^{n/2-2} 
    q(t^\prime(\eta')),
\label{2-9'}
\end{eqnarray}
where $p^2=(p^0)^2-|{\bf p}|^2$, and the contour of the integral 
of $p^0$ is specified in Figure~1 (see e.g., \cite{ref-5}).
We substitute $(\ref{2-9'})$ into $(\ref{1-7})$, then we obtain
\begin{equation}
 \left( \frac{\delta S_0}{\delta q}\right) \left(\tau \right)
  -(-H\eta(\tau))^{n/2-1} \int^0_{-\infty} d\eta^\prime 
  \int^\infty_{-\infty} \frac{d^n p}{(2\pi)^n}
    \frac{e^{ip^0(\eta-\eta^\prime )}}{p^2} 
     (-H\eta^\prime)^{n/2-2} q(\tau^\prime(\eta'))=0.
\label{1-10}
\end{equation}
After inserting the relation
\begin{eqnarray}
 q(\tau') =\frac{1}{2\pi}\int^\infty_{-\infty} d\omega'\mbox{ }
        \tilde{q}(\omega') e^{i\omega'\tau'},
\label{qomega}
\end{eqnarray}
into (\ref{1-10}), the Fourier transformation of equation (\ref{1-10})
yields
\begin{eqnarray}
   &&\widetilde{\left(\frac{\delta S_0}{\delta q}\right)}(\omega)
    -\int^0_{-\infty} d\eta \int^0_{-\infty} d\eta^\prime 
  \int^\infty_{-\infty}
       \frac{d^n p}{(2\pi)^n} \int^\infty_{-\infty}
  \frac{d\omega^\prime}{2\pi}
\nonumber
\\
   &&{\hspace{3cm}} \times
         (-H\eta)^{n/2-2} (-H\eta^\prime)^{n/2-2} 
       \frac{\widetilde{q}(\omega^\prime)}{p^2} 
        e^{i\left( \omega^\prime \tau^\prime -\omega\tau \right)}
        e^{ip^0 \left( \eta -\eta^\prime \right)}=0, 
\end{eqnarray}
where we defined
\begin{eqnarray}
  \widetilde{\left(\frac{\delta S_0}{\delta q}\right)}(\omega)
  = \int^\infty_{-\infty}d\tau e^{-i\omega\tau}
  \left( \frac{\delta S_0}{\delta q}\right) \left(\tau \right).
\label{Somega}
 \end{eqnarray}
With the use of the relation
\begin{eqnarray}
  \int^\infty_{-\infty}dp^0 
  {e^{ip^0 \left( \eta -\eta^\prime \right)}\over p^2}
  = \theta(\eta-\eta^\prime){i\pi\over|\bf p|} \left\{
       e^{i|\bf p|(\eta-\eta^\prime)}
      -e^{-i|\bf p|(\eta-\eta^\prime)} \right\},
\end{eqnarray}
we have
\begin{eqnarray}
  &&\widetilde{\left(\frac{\delta S_0}{\delta q}\right)}(\omega)
    -i\ \frac{\pi^{(n-1)/2}}{(2\pi)^n \Gamma\left( (n-1)/2 \right)}  
    \int^0_{-\infty} d\eta
    \int^0_{-\infty} d\eta^\prime 
  (-H\eta)^{n/2-2} (-H\eta^\prime)^{n/2-2} 
\nonumber
\\
  &&\hspace{1cm}  \times     \int^\infty_{-\infty} 
    d\omega^\prime\widetilde{q}(\omega^\prime) 
    e^{i\left( \omega^\prime \tau^\prime -\omega\tau \right)} 
     \theta(\eta-\eta^\prime) 
      \int^\infty_0 dp \mbox{ } p^{n-3} \left\{
         e^{ip(\eta-\eta^\prime +i\epsilon)}
           - e^{-ip(\eta-\eta^\prime -i\epsilon)} \right\}=0.
\end{eqnarray}
where we inserted $\pm i\epsilon$ for convergence of $p$-integration. 
Integration with respect to $p$ yields 
\begin{eqnarray}
  &&\widetilde{\left(\frac{\delta S_0}{\delta q}\right)}(\omega)
       -\frac{i\pi^{-(n+1)/2}}{2^n} \frac{\Gamma(n-2)}{\Gamma((n-1)/2)}
       \int^\infty_{-\infty}  \int^\infty_{-\infty}  d\tau d\tau^\prime 
   (-H\eta)^{(n-2)/2} (-H\eta^\prime)^{(n-2)/2}
\nonumber
\\
   &&\hspace{1cm} \times  
        \int^\infty_{-\infty} d\omega^\prime 
       \widetilde{q}(\omega^\prime) 
       e^{i(\omega^\prime \tau^\prime -\omega\tau)} \theta(\eta-\eta^\prime)
    \left\{ \frac{i^{2-n}}{\left( -\eta+\eta^\prime-i\epsilon\right)^{n-2}}
        -\frac{i^{2-n}}{\left( \eta-\eta^\prime+i\epsilon \right)^{n-2}}
           \right\}=0,
\end{eqnarray}
where we used the next relations,
\begin{eqnarray}
 && \int^\infty_0 \ x^n e^{-\alpha x}\ dx =\frac{\Gamma(n-1)}{\alpha^{n+1}}, 
\\
 && \Gamma(2z) = \frac{2^{2z}}{2\pi^{1/2}} \Gamma(z) \Gamma(z+1/2),
\end{eqnarray}
and $d\eta^\prime=(-H\eta^\prime)d\tau^\prime$. Then, by using the relation
\begin{eqnarray}
{(-H\eta)^{(n-2)/2}(-H\eta')^{(n-2)/2}\over (\eta-\eta'-i\epsilon)^{n-2}}
={H^{n-2}\over\left\{ 
      2\sinh{ \left(H (\tau-\tau')/2\right) -i\epsilon}\right\}^{n-2}},
\label{relationA}
\end{eqnarray}
and by introducing the variables
\begin{eqnarray}
 \Delta\tau \equiv \tau-\tau^\prime \ ,\ T \equiv \frac{\tau+\tau^\prime}2,
\label{T-vari}
\end{eqnarray}
we obtain the following equation after integrations
\begin{equation}
 \widetilde{\left( \frac{\delta S_0}{\delta q} \right) }(\omega)
  +{\mit{K}}(\omega) \tilde{q} (\omega)=0,
\label{1-11}
\end{equation}
where
\begin{eqnarray}
 {\mit{K}} (\omega)
   & = & \frac{i\Gamma(n/2-1)}{4\pi^{n/2}}
      {\left( \frac{H}{i}\right) }^{n-2}
\int^\infty_0 d(\Delta\tau) \mbox{ } e^{-i\omega \deltatau}
\nonumber \\
    &&  \quad 
 \times 
     \left[ \left\{ 2\sinh{\frac{H}{2}\deltatau -i\epsilon}\right\}^{2-n}
      -(-1)^n \left\{ 2\sinh{\frac{H}{2}\deltatau +i\epsilon} \right\}^{2-n}
        \right].
\label{1-12}
\end{eqnarray}
According to \cite{ref-1,ref-2}, the dissipative
coefficient $R (\omega)$ is obtained by
\begin{eqnarray}
  R(\omega)={-1\over \omega} {\rm Im}[K(\omega)].
\label{Romega}
\end{eqnarray}
After integration the dissipative coefficient is explicitly 
obtained from equation (\ref{1-12}) as
\begin{eqnarray}
 R(\omega) 
 & = & \frac{1}{2^{n-1}\pi^{(n-3)/2}} \frac{\omega^{n-4}}
  {\Gamma((n-1)/2)}
    \frac{\exp{\left( 2\pi\omega /H \right)}-1}
                        {\exp{\left( 2\pi\omega /H \right)}-(-1)^n}.
\label{1-13}
\end{eqnarray}

On the other hand, following the formula by Callen and Welton 
\cite{ref-3}, the FD relation for the thermal noise is 
written as
\begin{eqnarray}
 \langle \phi(x) \phi(x) \rangle 
   &\equiv& \int^\infty_0 dE \  \rho(E) f(E)
        \mbox{ } \langle E| \phi(x)\phi(x) |E\rangle 
    \nonumber \\
  & = & \frac2\pi \int^\infty_0 E(\omega,T) R(\omega) \mbox{ } d\omega, 
\label{1-14}
\end{eqnarray} 
where $\rho(E)$ is the density of states at the energy $E$ and 
$f(E)$ is the normalized Boltzmann factor, which satisfies
\begin{eqnarray}
 \frac{f\left( E+\omega\right)}{f\left( E \right) }
   =\exp{\left( -\frac{\omega}{T} \right)} 
\end{eqnarray}
with the temperature $T$, and $E(\omega,T)$ is defined by 
\begin{eqnarray}
  E(\omega,T)=\frac12 \omega 
   + \frac{\omega}{e^{\omega /T} -1}.
\end{eqnarray}
Here $E(\omega,T)$ can be regarded as the energy of a Bosonic 
harmonic oscillator with zero-point oscillation at the temperature $T$. 

We can easily show that the FD relation (\ref{1-14}) is satisfied
between the vacuum fluctuations and the dissipative coefficient.
The quantum field theory of a real scalar field in de Sitter 
spacetime has been well investigated \cite{BD}. In the case of 
the massless and conformally coupling, it is well-known that 
the two-point function is in proportion to that in the 
Minkowski spacetime. With the use of equation (\ref{relationA}),
the Wightman function leads to
\begin{equation}
 \langle 0_c| \phi(x(\tau)) \phi(x^\prime(\tau')) |0_c \rangle
  = \frac{\Gamma(n/2-1)}{4\pi^{n/2}} \left( \frac Hi \right)^{n-2}
     \left\{ 2\sinh{ \frac H2 \Delta\tau -i\epsilon }
        \right\}^{2-n} ,
\label{1-17}
\end{equation}
where $|0_c\rangle$ denotes the conformal vacuum. 
With the relation 
\begin{eqnarray*}
  \delta(\tau-\tau')=\int_{-\infty}^{\infty} 
  {d\omega \over 2\pi}e^{i\omega(\tau-\tau')}, 
\end{eqnarray*}
the fluctuation of the real scalar field is expressed as follows
\begin{eqnarray}
 \langle 0_c | \phi(x)^2 | 0_c\rangle
  & = & \int^\infty_{-\infty} d\tau^\prime 
        \int^\infty_{-\infty} \frac{d\omega}{2\pi}
     e^{i\omega (\tau-\tau^\prime)}
       \langle 0_c | \phi(x(\tau)) \phi(x'(\tau^\prime)) | 0_c\rangle
\nonumber \\
   & = & \frac{1}{2\pi} \int^\infty_{-\infty} F_n (\omega)\mbox{ } d\omega,
\label{1-15}
\end{eqnarray}
where the power spectrum, $F_n(\omega)$, is defined by
\begin{equation}
 F_n(\omega)=\int^\infty_{-\infty} d\tau^{\prime} 
       e^{-i\omega(\tau-\tau^\prime)}
     \langle 0_c | \phi(x(\tau)) \phi(x'(\tau^\prime)) | 0_c\rangle.
\label{1-16}
\end{equation}
Inserting equation $(\ref{1-17})$ into $(\ref{1-16})$ we obtain 
\begin{equation}
 F_n (\omega) =\frac{1}{2^{n-2}\pi^{(n-3)/2}}\   
     \frac{ \omega^{n-3}}{\Gamma ((n-1)/2) }
     \frac{1}{\exp{(2\pi\omega/H)}-(-1)^n }.
\label{1-18} 
\end{equation}
We finally conclude the existence of the FD relation
in $N$-dimensional de Sitter spacetime
\begin{equation}
 \langle 0_c | \phi(x(\tau))^2 | 0_c\rangle
  ={2\over \pi} \int_0^\infty E(\omega,H/2\pi) R(\omega) d\omega.
\label{1-19}
\end{equation}
From equation $(\ref{1-18})$ we can see that the statistical 
inversion occurs as in the case of the $N$-dimensional Rindler 
spacetime \cite{ref-1}. 
The FD relation (\ref{1-19}) holds irrespective of the statistical 
inversion.

\subsection{Dirac field}
Following the investigation by Terashima \cite{ref-1}, we next 
consider the system of a Dirac field $\psi$ and a detector
of a spinor $\Theta$ which is linearly coupled with the field.
The action of the system is 
\begin{eqnarray*}
 S=S_0(\Theta,\bar{\Theta})
      +S_{int}(\Theta,\bar{\Theta},\psi,\bar{\psi})
      +S_0(\psi,\bar{\psi}) ,
\end{eqnarray*}
where we defined
\begin{equation}
 S_0(\Theta,\bar{\Theta}) = \int d\tau 
      L(\Theta,\bar{\Theta},\dot{\Theta},\dot{\bar{\Theta}}),
\label{1-20}
\end{equation} 
\begin{equation}
 S_0(\psi,\bar{\psi}) = \int d^nx (\det{V}) \left[ \  \frac{i}{2}
    \left\{ \bar{\psi} \gamma^\mu \nabla_\mu \psi 
     -(\nabla_\mu \bar{\psi})\gamma^\mu \psi \right\} 
     -m\bar{\psi}\psi \right],
\label{1-21}
\end{equation}
\begin{eqnarray}
 S_{int}(\Theta,\bar{\Theta},\psi,\bar{\psi}) 
   & = & \int d\tau \left\{  \bar{\Theta}(\tau) \psi(x(\tau)) 
               +\bar{\psi}(x(\tau)) \Theta(\tau) \right\}
\nonumber  \\
       & = & \int d\tau d^nx \left( \bar{\Theta}(\tau) \psi(x) 
             +\bar{\psi}(x) \Theta(\tau)\right) 
          \delta^n (x -x_0(\tau)),
\label{1-22}
\end{eqnarray}
where $\Theta(\tau)$ and $\bar{\Theta}(\tau)(=\Theta(\tau)^\dagger\gamma^0)$ are 
Grassmann numbers, and $\nabla_\mu$ is defined by 
\begin{equation}
 \nabla_\mu \equiv \partial_\mu + \Gamma_\mu ,\qquad
  \Gamma_\mu = \frac12 \Sigma^{\alpha \beta} {V_{\alpha}}^\nu (x)
     (\nabla_\mu V_{\beta \nu}),
\end{equation}
with the vierbein $V_{\alpha}^{~\nu}$ and the generator of Lorentz 
transformation $\Sigma^{\alpha \beta}$ \cite{ref-5}. Equations of the motion
are written 
\begin{eqnarray}
\left( \frac{\delta S_0}{\delta \bar{\Theta}} \right) (\tau) 
   +\int d^n x \psi (x) 
      \delta^n (x-x_0(\tau)) =0,
\label{1-24} \\
 (i\snab -m)\psi 
   =\frac{-1}{\det{V}}\int d\tau \Theta(\tau) \delta^n (x-x(\tau)),
\label{1-23} 
\end{eqnarray}
where, as in the case of the scalar field, we defined 
\begin{eqnarray*}
 \left(\frac{\delta S_0}{\delta \bar{\Theta}}\right)(\tau) \equiv
  -\frac{d}{d\tau} \left( \frac{\partial L}{\partial \dot{\bar{\Theta}}} \right)
     +\frac{\partial L}{\partial \bar{\Theta}} ~.
\end{eqnarray*}
For simplicity we restrict $\psi$ is the massless field in this 
section. Similar to the case of the scalar field, we introduce 
the conformal field $\psi_c$ defined with a  conformal weight as 
\begin{equation}
 \psi(x) = \left( -H\eta \right)^{(n-1)/2} \psi_c (x). 
\end{equation}
In terms of $\psi_c$, equations (\ref{1-24}) and (\ref{1-23})
reduce to
\begin{eqnarray}
 \left( \frac{\delta S_0}{\delta \bar{\Theta}} \right) (\tau) 
   + (-H \eta(\tau))^{(n-1)/2} \psi_c (x_0(\tau)) =0 , 
\label{1-23'}
\end{eqnarray}
and
\begin{eqnarray}
i\spar \psi_c(x) 
    =  -(-H\eta)^{(n-3)/2}
     \Theta(\tau(\eta)) \delta^{n-1}(\bfx-\bfx_0) ,
\label{1-23''}
\end{eqnarray}
respectively, where we defined 
$\spar \equiv \gamma^0 \partial_\eta +\gamma^i \partial_i$
with $(i=1,2,\ldots,n-1)$, and assumed the detector at rest of the
spatial coordinate $\bfx=\bfx_0$. 
As is done in the previous subsection, equation (\ref{1-23''}) can be 
solved by using the retarded Green function, which we denote by
${\bar S}_R (x,x^\prime)$, as follows,
\begin{equation}
  \psi_c(x)  =  \int^0_{-\infty} d\eta^\prime 
        \int^\infty_{-\infty} d^{n-1}x^\prime 
        {\bar S}_R(x,x^\prime) \left( -H\eta^\prime \right)^{(n-3)/2}
         \Theta(\tau^\prime(\eta')) \delta^{n-1}({\bf x}^\prime-{\bf x}_0), 
\label{psic}
\end{equation}
where we assume that the retarded Green function satisfies
\begin{equation}
 i\spar_x {\bar S}_R (x, x^\prime)
       =-\delta^n (x-x^\prime),
\end{equation}  
and the boundary condition $\bar S_R (x,x^\prime)=0$ for 
$\eta - \eta^\prime < 0$.
Substituting equation (\ref{psic}) into $(\ref{1-23'})$, we have
\begin{eqnarray}
 \left(\frac{\delta S_0}{\delta \bar{\Theta}} \right)(\tau)
  & = & -\int^\infty_{-\infty} d\tau^\prime  
      \bar S_R(x_0(\tau),x_0(\tau^\prime))
     \left( -H\eta^\prime(\tau^\prime) \right)^{(n-1)/2}
      \left( -H\eta(\tau) \right)^{(n-1)/2} \Theta(\tau^\prime). 
\label{deltasdq}
\end{eqnarray}
The Fourier transformation of equation (\ref{deltasdq}) yields
\begin{eqnarray}
 \widetilde{\left( \frac{\delta S_0}{\delta \bar{\Theta}}\right)}(\omega)
  &=& -\int^\infty_{-\infty} d\tau^\prime 
  \int^\infty_{-\infty} d\tau \int^\infty_{-\infty}
  \frac{d\omega^\prime}{2\pi} {\tilde \Theta}(\omega^\prime)
       e^{i(\omega^\prime \tau^\prime - \omega\tau )} 
\nonumber   
 \\&& \qquad \times
   \bar S_R(x_0(\tau),x_0(\tau^\prime)) 
     \left( -H\eta^\prime(\tau^\prime) \right)^{(n-1)/2}
     \left( -H\eta(\tau) \right)^{(n-1)/2} ,
\label{DiracA}
\end{eqnarray}
where the Fourier coefficients ${\tilde \Theta}(\omega^\prime)$ and 
$(\widetilde{\delta S_0}/\delta \bar{\Theta})(\omega)$ are
defined in similar ways to (\ref{qomega}) and (\ref{Somega}),
respectively. 
The massless Green function on the flat spacetime is well-known, and
we can write down $\bar S_R(x,x')$ in an explicit form. With the use of 
the relation (\ref{relationA}), we can write
\begin{eqnarray}
  &&\bar S_R(x_0(\tau),x_0(\tau^\prime)) 
  \left( -H\eta(\tau)\right)^{(n-1)/2} 
  \left( -H\eta^\prime(\tau')\right)^{(n-1)/2}
   = i  \gamma^0 \frac{\Gamma(n/2)}{2\pi^{n/2}}
             \left( \frac{H}{i} \right)^{n-1}     
\nonumber\\
 &&\hspace{1.5cm}  \times \theta (\Delta\tau)
  \left[ \left\{ 2\sinh{\frac H2 \Delta\tau -i\epsilon} \right\}^{1-n}
     -(-1)^n \left\{ 2\sinh{\frac H2 \Delta\tau +i\epsilon} \right\}^{1-n} 
        \right].
\end{eqnarray} 
Keeping this relation in mind, after integration using 
the variables $T$ and $\Delta\tau$ defined by equation 
$(\ref{T-vari})$, equation (\ref{DiracA}) yields 
\begin{eqnarray}
 \widetilde{\left( \frac{\delta S_0}{\delta \bar{\Theta}}\right)}(\omega)
  &=& -\int d(\Delta\tau) \bar S_R(x_0(\tau),x_0(\tau^\prime))
    \left( -H\eta^\prime \right)^{(n-1)/2} \left( -H\eta \right)^{(n-1)/2}
      e^{-i\omega \Delta\tau} {\tilde \Theta}(\omega).
\end{eqnarray}
Then we obtain the effective equation of motion for the Fourier coefficient
\begin{equation}
 \widetilde{\left( \frac{\delta S_0}{\delta \bar{\Theta}} \right)} (\omega)
      +{\mit K}_{1/2} (\omega) \widetilde{\Theta} (\omega)  = 0,
\label{1-25}
\end{equation}
where 
\begin{eqnarray}
 {\mit K}_{1/2} (\omega) &=& i\gamma^0 
      \frac{\Gamma (n/2)}{2\pi^{n/2}}
     \left( \frac{H}{i}\right)^{n-1} 
       \int^\infty_0 d(\Delta\tau) e^{-i\omega\Delta\tau}
\nonumber  \\   && \qquad \times 
           \left[\left\{ 2\sinh{\frac{H}{2}
                     \Delta\tau -i\epsilon}\right\}^{1-n}  
             -(-1)^n \left\{ 2\sinh{\frac{H}{2}
                \    \Delta\tau+i\epsilon}\right\}^{1-n} \right].
\label{1-26}
\end{eqnarray}
In the similar way to equation (\ref{Romega}), the dissipative 
coefficient is defined (c.f., refs.\cite{ref-1,ref-2}),
\begin{eqnarray}
  R_{1/2} (\omega)={-1\over \omega }{\rm Im}{[{\mit K}_{1/2} (\omega)]}.
\label{defRF}
\end{eqnarray}
After integration, we obtain
\begin{equation}
 R_{1/2}(\omega)
   = -\gamma^0 \  \frac{\omega^{n-3}}{2^{n-1}\pi^{(n-3)/2}}
     \  \frac{1}{\Gamma((n-1)/2)}
     \  \frac{\exp{\left( 2\pi\omega/H \right)}+1}
            {\exp{\left( 2\pi\omega/H  \right)}+(-1)^n}.
\label{1-27}
\end{equation}

A fermionic version of the FD relation is presented in \cite{ref-2},
which is applied to the system of the Dirac field and the detector
in the Rindler spacetime in \cite{ref-1}. The fermionic FD relation
is written 
\begin{equation}
 {\rm Tr} \left[ \gamma_0 \langle \psi(x) \bar{\psi} (x) \rangle \right]
   =  \frac{1}{\pi} \int^\infty_{-\infty} d\omega \mbox{ }
        \frac{-\omega}{e^{-\omega/T} +1} \mbox{ } 
           {\rm Tr} \left[ \gamma_0 R_{1/2}(\omega)\right].
\label{1-28}
\end{equation}
If the dissipative coefficient satisfies  
$R_{1/2}(-\omega)= -R_{1/2}(\omega)$,
equation $(\ref{1-28})$ can be rewritten
\begin{equation}
 {\rm Tr} \left[ \gamma_0 \langle \psi(x) \bar{\psi} (x) \rangle \right] 
   =  -{1\over \pi}  \int^\infty_0 \omega 
      {\rm Tr}\left[ \gamma_0R_{1/2}(\omega) \right]  d\omega.
\label{1-28b}
\end{equation}

On the other hand, quantum field theory of the Dirac field in de 
Sitter spacetime is well understood. In the case of the massless 
field, the Wightman function in the $N$-dimensional de Sitter 
spacetime is written\cite{ref-4,ref-5}
\begin{equation}
 \langle 0| \psi(x) \bar{\psi} (x^\prime) |0 \rangle 
  = \gamma^0 \frac{\Gamma(n/2)}{2\pi^{n/2}} \left(\frac Hi \right)^{n-1}
    \left\{ 2\sinh{\frac H2} \Delta\tau -i\epsilon  \right\}^{1-n}.
\label{1-28'}    
\end{equation}
Similar to the case of the scalar field, we write the 
fluctuation of the Dirac field as
\begin{eqnarray}
 {\rm Tr} \left[ \gamma_0\langle 0| \psi(x) \bar{\psi} (x) |0 \rangle \right]
 &=&  \int^\infty_{-\infty} d\tau^\prime 
        \int^\infty_{-\infty} \frac{d\omega}{2\pi} \mbox{ }
                  e^{-i\omega (\tau -\tau^\prime)} 
      {\rm Tr} \left[\gamma_0 \langle 0 | \psi(x) \bar{\psi} (x^\prime)
        | 0 \rangle \right]
\nonumber \\
 &\equiv& \frac{\Delta_n}{2\pi} \int^\infty_{-\infty} 
      F_{1/2}(\omega) \mbox{ } d\omega ,
\label{1-28''}
\end{eqnarray}
where $\Delta_n$ is the dimensions of the $\gamma$-matrices, and 
the power spectrum is defined as a scalar quantity by
\begin{equation}
 F_{1/2}(\omega) \equiv \frac 1{\Delta_n} 
     \int^\infty_{-\infty} d\tau^\prime 
                  e^{-i\omega (\tau -\tau^\prime)} 
     {\rm Tr}\left[\gamma_0 \langle 0 | \psi(x) 
          \bar{\psi} (x^\prime) | 0 \rangle \right].
\label{1-28'''}
\end{equation}
Inserting equation $(\ref{1-28'})$ into $(\ref{1-28'''})$, we obtain 
\begin{equation}
 F_{1/2}(\omega) = \  \frac{\omega^{n-2}}{2^{n-2}\pi^{(n-1)/2}}
              \  \frac{1}{\Gamma ((n-1)/2)}
          \  \frac{1}{\exp{\left( 2\pi\omega/H \right)+(-1)^n }}.
\label{1-29}
\end{equation}
This means the existence of the FD relation (\ref{1-28}) or 
(\ref{1-28b}) with the dissipative coefficient (\ref{1-27}). 
From equation $(\ref{1-29})$ we can see that the statistical 
inversion occurs in the fermionic case too, however, the 
FD-relation exists irrespective of the statistical inversion.

\section{Essential conditions for the FD relation}
\subsection{real scalar field}

In the previous section we have derived the FD relation
in an explicit manner, according to the method proposed by 
Terashima \cite{ref-1}. In the previous section, however,
we have considered the simple cases of massless fields in de 
Sitter spacetime.
In this section we summarize the calculations in the previous 
section from a general point of view, which clarifies why the 
Terashima's prescription works properly. Let us start 
from considering the classical equations of 
motion of the scalar field and the detector (\ref{1-4}) and
(\ref{1-5}).
As we have done in the previous section, the field equation 
(\ref{1-5}) is formally solved by introducing the retarded 
Green function, which satisfies
\begin{equation}
 \left( \nabla^\mu_x \nabla_\mu +m^2 +\xi R \right) G_R (x,x^\prime)
   = \frac{1}{\sqrt{-g}} \mbox{ }\delta^n (x-x^\prime),
\label{2-3}
\end{equation}
to give
\begin{equation}
  \phi(x)=\int d^n x'G_R(x,x') \int d\tau q(\tau) \delta^n(x'-x_0(\tau)).
\label{2-3'}
\end{equation}
Inserting this into equation (\ref{1-4}), we have
\begin{eqnarray}
 &&\left( \frac{\delta S_0}{\delta q}\right) \left(\tau \right)
   + \int d\tau' G_R(x_0(\tau),x_0'(\tau')) q(\tau') = 0.
\label{deltas}
\end{eqnarray}
Fourier transformation of this equation (\ref{deltas}) yields 
the same expression as (\ref{1-11})
but with
\begin{eqnarray}
 &&{ \mit{K}} (\omega) ={1\over 2\pi}
  \int^\infty_{-\infty} d\tau 
  \int^\infty_{-\infty} d\tau'
  \int^\infty_{-\infty} d\omega' \mbox{ } 
          G_R (x(\tau),x'(\tau')) e^{-i(\omega \tau-\omega'\tau')},
\label{2-5}
\end{eqnarray}
where the Fourier coefficients ${\tilde q}(\omega^\prime)$ and 
$(\widetilde{\delta S_0}/\delta {q})(\omega)$ are
defined by (\ref{qomega}) and (\ref{Somega}), respectively. 
In the case that the retarded Green function is the function 
of $\Delta \tau(=\tau-\tau')$, i.e., 
$G_R (x(\tau),x'(\tau'))=G_R(\Delta \tau)$, 
equation (\ref{2-5}) reduces to 
\begin{eqnarray}
 && \mit{K} (\omega)  =\int^\infty_{-\infty} d(\Delta\tau)  \mbox{ } 
          G_R (\Delta\tau) e^{-i\omega \Delta\tau}.
\label{2-5'}
\end{eqnarray}
The dissipative coefficient defined by (\ref{Romega}) becomes
\begin{eqnarray}
 &&R(\omega) = -{1\over\omega} {\rm Im}\biggl[ 
          \int^\infty_{-\infty} d(\Delta\tau) G_R(\Delta\tau)
           \mbox{ }   e^{-i\omega \Delta\tau}\mbox{ }  \biggr].
\label{2-6}
\end{eqnarray}

In the meanwhile, it is well-known that, in the thermal field theory,
the Wightman functions satisfy a periodicity condition in 
the direction of imaginary 
time, which is called the Kubo-Martin-Schwinger (KMS) condition 
\cite{Kubo,MS}. 
In the quantum field theory in curved spacetime, we assume that 
the Wightman functions, which are defined by 
\begin{eqnarray}
 && G^+ (\Delta\tau) \equiv 
     \langle 0| \phi(x(\tau)) 
        \phi(x^{\prime}(\tau^\prime)) |0 \rangle  ~,\\
 && G^-(\Delta\tau) \equiv 
     \langle 0| \phi(x^{\prime}(\tau^\prime))     
            \phi(x(\tau)) |0 \rangle      ~,
\end{eqnarray}
satisfy the KMS condition on the complex 
plane of the detector's proper time,
\begin{equation}
 G^+ (\Delta\tau -i\beta) = G^- (\Delta\tau),
\label{2-7}
\end{equation} 
where $\beta$ is a period of the periodicity in the direction 
of imaginary time. In the thermal field theory, $\beta=1/T$.
For the massless (conformally coupled) scalar field, assuming a 
detector at rest in de Sitter spacetime, the Wightman functions are
\begin{eqnarray}
  G^\pm (\Delta\tau)  ={\Gamma(n/2-1)\over 4\pi^{n/2}}{H^{n-2}\over 
  (\pm i)^{n-2}}
 \Biggl[{1\over2 \sinh(H\Delta\tau/2)\mp i\epsilon}\Biggr]^{n-2} ,
\end{eqnarray}
hence we can easily check that the Wightman functions of the conformal 
vacuum state satisfy the KMS condition with $\beta=2\pi/H$.

Following the investigations by Takagi \cite{ref-4} and Ooguri 
\cite{Ooguri}, let us introduce the Fourier coefficients, which 
are defined by
\def\Dtau{{\Delta\tau}}
\begin{eqnarray}
 && F^\pm (\omega) \equiv \int^\infty_{-\infty} 
           G^\pm(\Dtau) e^{-i\omega \Dtau}\mbox{ } d(\Dtau) .
\label{2-9}
\end{eqnarray}
Because of the definition, the Wightman functions are related to 
each other as $G^+(-\Dtau)=G^-(\Dtau)$.
This relation is equivalent to $F^+(-\omega)=F^-(\omega)$.
On the other hand, the result of the KMS condition (\ref{2-7})
for the Fourier coefficients is derived as follows: 
By extending $\Dtau$-integration in (\ref{2-9}) to the complex plane,
we can deform the path of the integration from $C_1$ on the real
axis to $C_2$ on the complex plane (see Figure 2),
\begin{eqnarray}
  F^+ (\omega) &=& \int_{C_2} G^+(z) e^{-i\omega z} dz.
\label{F+A}
\end{eqnarray}
This can be done unless a singular pole appears 
in the region between the paths $C_1$ and $C_2$ in Figure 2.
Then, using the KMS condition, (\ref{F+A}) is written as
\cite{ref-4,Ooguri},
\begin{eqnarray}
 F^+ (\omega)=e^{-\beta\omega}\int^\infty_{-\infty} 
 G^-(\Dtau) e^{-i\omega\Dtau} d(\Dtau)
 =e^{-\beta\omega} F^-(\omega).
\label{2-10}
\end{eqnarray}

The retarded Green function is related with the Wightman functions
(e.g., \cite{ref-5}),
\begin{eqnarray}
  G_R (x,x') =i\theta(t-t') \left( G^+(x,x') - G^-(x,x') \right).
\label{GGG}
\end{eqnarray}
Inserting this relation into equation (\ref{2-6}), we obtain the
dissipative coefficient 
\begin{equation}
  R(\omega) =\frac{e^{\beta\omega} -1}{2\omega}F^+(\omega),
\label{2-11}  
\end{equation}
where we used (\ref{2-10}) and $G^+(-\Dtau)=G^-(\Dtau)$, 
which is equivalent to $F^+(-\omega)=F^-(\omega)$.
Finally by using equation (\ref{2-11}) the FD relation
is derived as an identity 
\begin{eqnarray}
         \langle 0| \phi(x)^2 |0 \rangle
   \equiv  {1\over 2\pi}\int^\infty_{-\infty} F^+(\omega)\mbox{ }d\omega
     =  {2\over \pi} \int^\infty_0
        E(\omega,1/\beta) R(\omega) \mbox{ } d\omega.
\label{2-13}
\end{eqnarray}

\subsection{Massive scalar field}
Following the above argument, the derivation of the 
FD relation can be generalized to the real scalar field with 
any mass $m$ in de Sitter spacetime. 
The Wightman functions for the massive field in 
$N$-dimensional de Sitter spacetime is
\begin{eqnarray}
 &&G^\pm(x,x^\prime) = {H^{n-2}\over(4\pi)^{n/2}}
           {\Gamma(a_+)\Gamma(a_-)\over\Gamma(c)}   
           {}_2F_1(a_+,a_-,c;1+z_\pm),
\label{4-1}
\end{eqnarray} 
where
\begin{eqnarray}
 &&a_\pm = {1\over 2}\left( n-1 \pm \sqrt{(n-1)^2-4m^2/H^2}\right),
\hspace{1cm} c={n\over 2},
\end{eqnarray}
and $z_\pm$ is in proportion to the geodesic distance 
between the two-points $x$ and $x'$, which is written 
using the coordinate of the spatially flat chart
\begin{eqnarray}
  z_\pm={(\eta-\eta' \mp i\epsilon)^2-|\bfx-\bfx'|^2\over 4 \eta\eta'}.
\label{defgeodesic}
\end{eqnarray} 
For the two-points on a geodesics of the detector, $x(\tau)$ and $x'(\tau')$,
we have
\begin{eqnarray}
  z_\pm=\left( \sinh{\frac H2 \Delta\tau} \mp i\epsilon \right)^2. 
\label{defzpm}
\end{eqnarray} 
Hence it is apparent that the Wightman functions satisfy the 
periodicity condition (\ref{2-7}). 

A delicate condition in deriving the FD relation might be the structure 
of singular poles of the Wightman functions on the complex plane of the 
imaginary time. If $G^+(\Dtau)$ has additional singular poles 
on the region between the contours $C_1$ and $C_2$ (see Fig.~2), 
equation (\ref{2-10} or \ref{2-21}) can not be satisfied. 
For a physical situation, it would be expected that the
singular pole of $G^{+}(x,x')$ appears only at $x=x'$. Hence 
the condition for the singular poles of the Wightman functions
would be satisfied. Actually in the case of the de Sitter spacetime, 
the Wightman functions satisfy this condition because of a 
property of the hypergeometric function in (\ref{4-1}).
Thus the FD relation holds between the dissipative coefficient of 
the detector's motion and the vacuum noise of the field with any mass 
as long as the detector moves along a geodesics.

\subsection{Dirac field}
We next consider the case of the Dirac field. We show that 
the argument similar to the case of the scalar field holds. 
Let us start with equations of motion (\ref{1-24}) and  (\ref{1-23}).
Equation (\ref{1-24}) is solved by introducing the retarded Green
function, which satisfies equation
\begin{eqnarray}
 \left(i\snab_x -m \right) S_R(x,x^\prime)
     =\frac{-1}{\det{V}} \delta^n (x-x^\prime),
\end{eqnarray}
as
\begin{eqnarray}
  \psi(x)=\int d^n x'S_R(x,x') \int d\tau \Theta(\tau) \delta^n(x'-x_0(\tau)).
\end{eqnarray}
Inserting this expression into equation (\ref{1-24}), we have
\begin{eqnarray}
  {\left(\frac{\delta S_0}{\delta\bar{\Theta}}\right)}(\tau)
  +\int d\tau' S_R(x_0(\tau),x_0'(\tau')) \Theta(\tau') = 0.
\end{eqnarray}
In the similar way to the case of the scalar field, we assume
that the retarded Green function is a function of only $\Dtau$, 
i.e., $S_R(x_0(\tau),x_0'(\tau'))=S_R(\Dtau)$. In this case
the Fourier transformation of this equation yields the same 
expression as (\ref{1-25}) with
\begin{eqnarray}
 && {\mit K}_{1/2} (\omega) =\int_{-\infty}^\infty 
  d(\Dtau) \mbox{ } S_R (\Dtau) 
  e^{-i\omega\Dtau}.
\label{2-17} 
\end{eqnarray}
Hence the dissipative coefficient defined by (\ref{defRF}) is
\begin{eqnarray}
  R_{1/2}(\omega)=-{1\over \omega} {\rm Im}
  \biggl[\int_{-\infty}^\infty  d(\Dtau) \mbox{ } S_R (\Dtau) 
  e^{-i\omega\Dtau}\biggr].
\label{disspF}
\end{eqnarray}

For the Dirac field we define
\begin{eqnarray}
 && S^+ (\Delta\tau) \equiv 
   \langle 0| \psi(x(\tau)) \bar{\psi}
         (x^{\prime}(\tau^\prime)) |0\rangle , \\
 && S^- (\Delta\tau) \equiv 
   \langle 0| \bar{\psi} (x^{\prime}(\tau^\prime))
        \psi (x(\tau)) |0\rangle,
\end{eqnarray}
and assume that these functions satisfy the KMS condition,
\begin{equation}
 S^+ (\Delta\tau -i\beta) = S^- (\Delta\tau) .
\label{2-18}  
\end{equation}
By the definition of the functions $S^\pm(\Delta\tau)$, we have
\begin{equation}
 S^- (\Delta\tau)  = S^+(-\Delta\tau)|_{m \to -m},
\label{SS}
\end{equation}
where we used the relation
$ S^\pm (x,x^\prime) =\pm \left( i\snab_x +m \right) G^\pm (x,x^\prime)$
and the assumption of the detector being at rest.
The Fourier coefficients are defined by 
\begin{equation}
 F^\pm_{1/2} (\omega) = \frac 1{\Delta_n} {\rm Tr} \left[ \gamma_0
        \int^\infty_{-\infty} S^\pm (\Delta \tau) e^{-i \omega\Delta\tau}
            d(\Delta\tau) \right].
\label{3+2}
\end{equation}
Then we have $ F^+_{1/2}(-\omega)=F^-_{1/2}(\omega)$ from (\ref{SS}).
We repeat the argument to derive (\ref{2-10}), and obtain
\begin{equation}
 F^+_{1/2} (\omega) = e^{-\omega\beta} F^-_{1/2} (\omega).
\label{2-21}
\end{equation}

For the Dirac field the retarded Green function should 
be related with $S^{\pm}$ as
\begin{eqnarray}
 S_R(x,x^\prime) = i\theta(t-t^\prime) 
         \left\{ S^+ (x,x^\prime)+ S^- (x,x^\prime) \right\}.
\label{SSS}
\end{eqnarray}
Inserting this relation into (\ref{disspF}), we have the 
dissipative coefficient
\begin{equation}
 {1\over \Delta_n}{\rm Tr}\bigl[ \gamma_0 R_{1/2}(\omega)\bigr] = -\  
  \frac{1+ e^{\beta\omega}}{2\omega}F^+_{1/2} (\omega).
\label{2-23}
\end{equation}
This implies the existence of the FD relation, 
\begin{eqnarray}
  && {\rm Tr}\bigl[ \gamma_0 \langle 0| \psi(x) \bar{\psi} (x) |0 \rangle
  \bigr] = {\Delta_n\over 2\pi} \int^\infty_{-\infty} 
  F^+_{1/2} (\omega) \mbox{ } d\omega
  =-{1\over \pi} \int^\infty_0\omega 
  {\rm Tr}\bigl[ \gamma_0 R_{1/2}(\omega)\bigr] \mbox{ } d\omega.
\label{2-24}  
\end{eqnarray}

\section{Summary and Discussions}
In summary we have investigated the FD relation between the 
dissipative coefficients of a detector and the vacuum noise 
of the scalar and the Dirac fields which are linearly coupled to 
each other in the $N$-dimensional de Sitter spacetime. 
We have derived the dissipative coefficient from the classical 
equations of motion, then the existence of the FD relation 
has been shown in an explicit manner by inserting the
dissipative coefficient into the FD relation of 
Callen and Welton \cite{ref-3}. 
Though our result might depend on the model of the couplings between 
the detector and fields, this investigation is interesting because 
information of quantum fluctuations can be obtained by observing 
the classical motion of the detector in principle.
As is shown in $\S$ 3, the periodicity in the Wightman functions 
(the KMS condition) is essential for the existence of the FD 
relation. Hence the FD relation exists irrelevant to the statistical 
inversion on the contrary to naive expectation associated with the
inversion phenomenon.

Some equations in the present paper are well-recognized so far. One 
is the fact that the FD relation of the vacuum noise in de Sitter 
spacetime has been well-known in connection with the the KMS condition 
of the Wightman functions (see e.g., \cite{ref-4} and references 
therein). However, our explicit derivation of the FD relation 
according to the Terashima prescription is instructive to 
understand why the Terashima's method works properly and 
might give a hint as to the origin of the statistical inversion 
phenomenon in odd dimensions of de Sitter spacetime. Our 
investigation is useful to show the key conditions so that the 
Terashima's prescription works properly in general curved spacetime.
Our calculation also demonstrates the validity of the use of the 
retarded Green function to obtain a dissipative coefficient 
from classical equations of motion.

It is apparent that the vacuum state of the field in curved 
spacetime and the boundary condition of the retarded Green 
function should be carefully chosen for the FD relation existing.
For example if we adopt a vacuum state in de Sitter spacetime other 
than the Bunch-Davies vacuum, the existence of the FD relation is 
not guaranteed. Furthermore it should be noted that we assumed 
that the retarded Green function is related with the Wightman functions 
by equation (\ref{GGG}) or (\ref{SSS}) that depend on the choice of 
the vacuum state.

Finally we mention a connection of the FD relation with the Euclidean 
path integral approach. 
It has been pointed out that the thermal property of the Hawking 
radiation is traced back to the periodicity in time of the 
Euclidean section of the black hole spacetime \cite{ref-7}.
That is, if a curved spacetime possesses the periodicity in its 
Euclidean section, then the thermal property of a field on the 
Lorentzian spacetime can be expected.
The Bunch-Davies vacuum corresponds to the Euclidean vacuum of the 
de Sitter spacetime, hence the periodicity of the Wightman 
functions can be regarded as the result of the periodicity of 
Euclidean section of the de Sitter spacetime.
As long as the retarded Green function relevant to the 
Euclidean vacuum is adopted, the prescription proposed by 
Terashima is supposed to work properly in the black hole spacetime.

\acknowledgments
We would like to thank Misao Sasaki, Takahiro Tanaka, Masafumi Seriu, 
and Yasufumi Kojima for useful conversations on this topics. 
This research was supported by the Inamori Foundation and by the 
Grants-in-Aid by the Ministry of Education, Science, Sports
and Culture of Japan (11640280).

\newpage

\pagestyle{empty}
\begin{figure}[t]
\centerline{\epsfxsize=15cm \epsffile{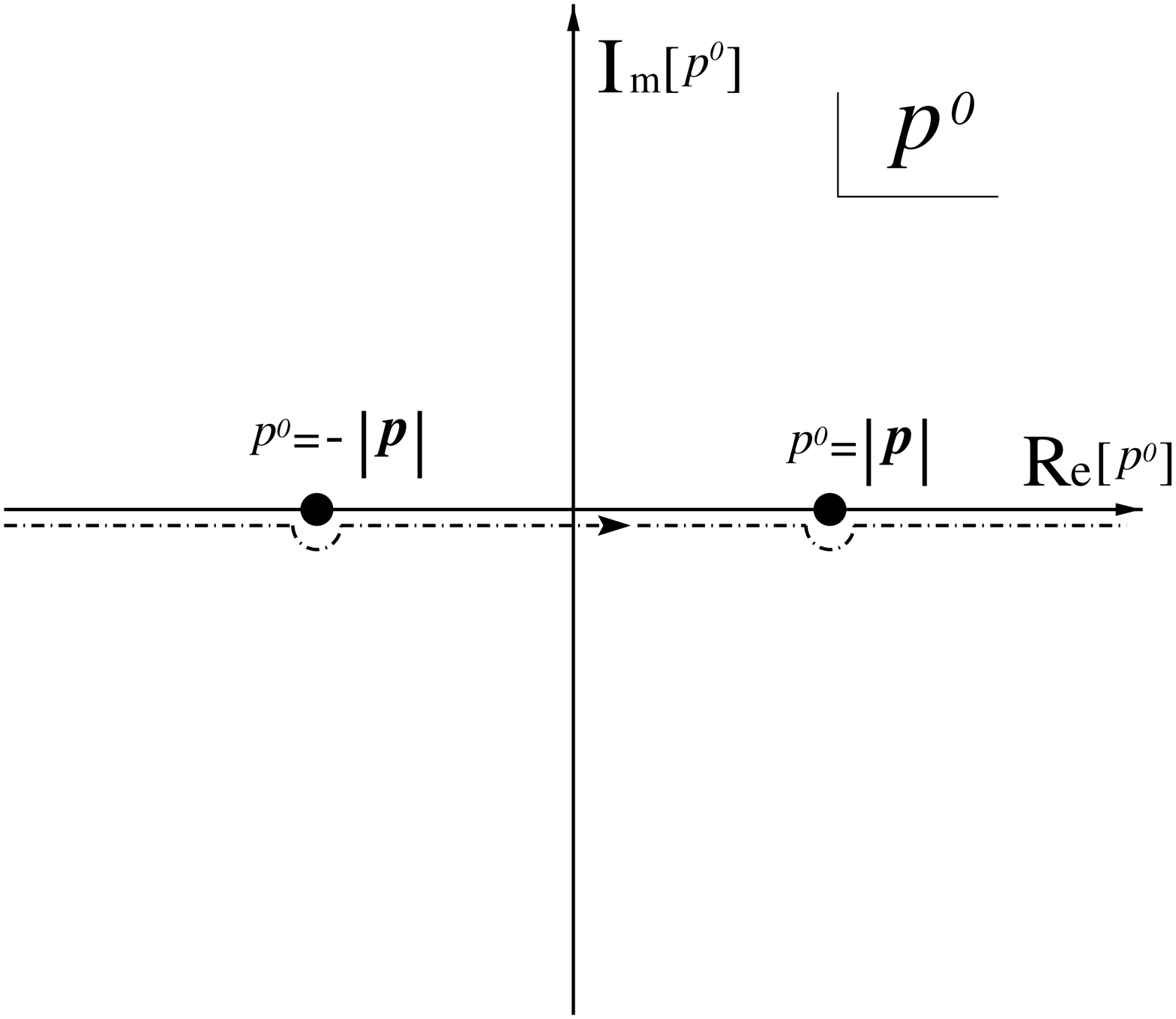}}
\caption{Contour of integration on the complex plane of $p^0$ 
 to obtain the retarded Green function. \label{fig1}
}
\end{figure}
\newpage
\begin{figure}[t]
\centerline{\epsfxsize=15cm \epsffile{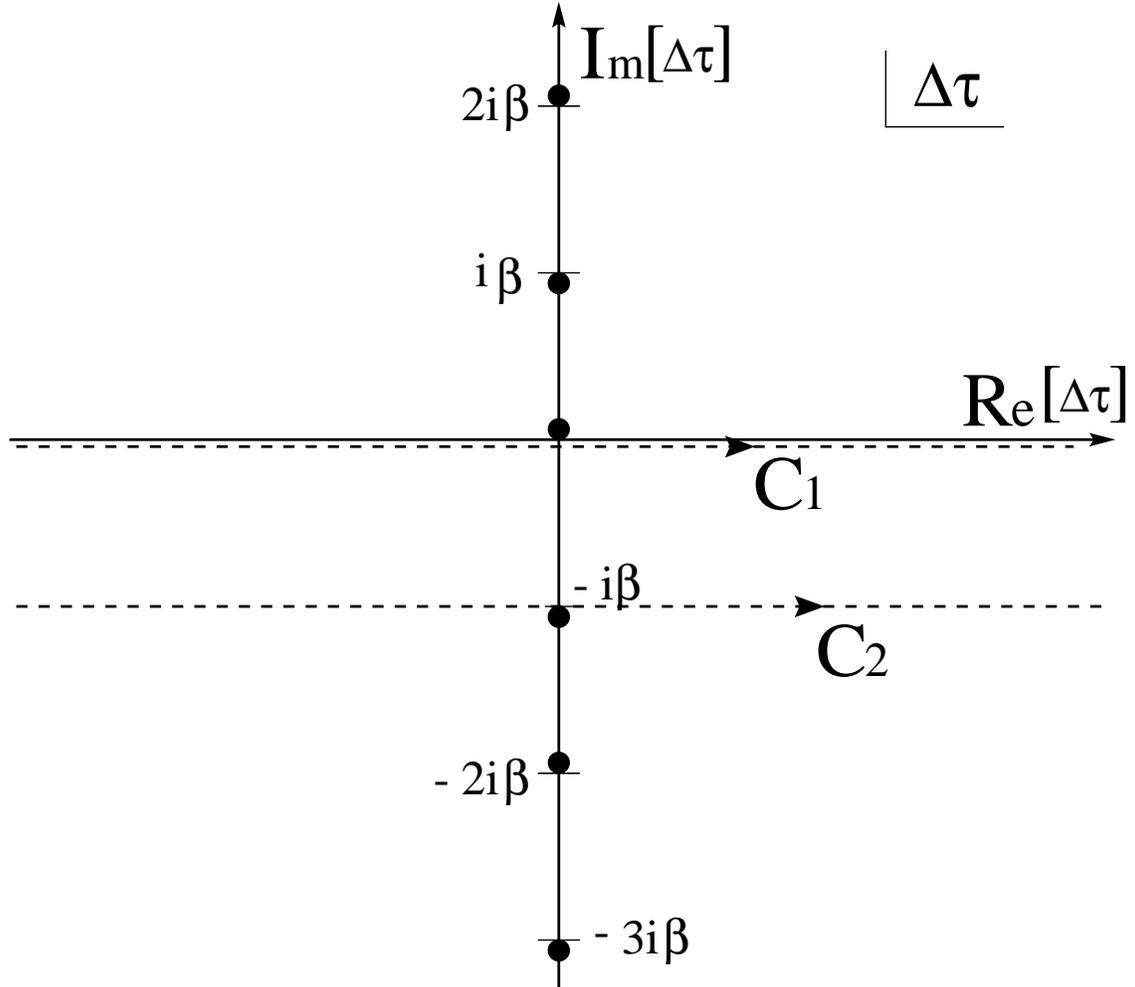}}
\caption{Structure of $G^+(\Dtau)$ on the complex plane of $\Dtau$. 
The black circles show the positions of the singular poles of
$G^+(\Dtau)$. $C_1$ denotes the contour of the integration in 
(3.11), and $C_2$ denotes the contour in (3.12). \label{fig2}
 }
\end{figure}

\end{document}